\documentclass[aps,eqsecnum,preprint,preprintnumbers,12pt,amsfonts]{revtex4}
\usepackage{epsfig,psfrag}
\usepackage{latexsym}

\newcommand{\be}{\begin{equation}}
\newcommand{\ee}{\end{equation}}
\newcommand{\ba}{\begin{eqnarray}}
\newcommand{\ea}{\end{eqnarray}}
\def\nn{\nonumber}

\begin{document}

\title{QCD in the Spatial Axial Gauge}

\author{Marek Kras\v nansk\'y\footnote{mkras@phys.uconn.edu}}

\affiliation{Department of Physics,
University of Connecticut,
Storrs, CT 06269-3046, USA}

\begin{abstract}

  Canonical quantization of a gauge theory in the spatial axial gauge produces an anisotropic Hamiltonian and matter particles
  surrounded by physically unrealistic asymmetric electric or chromoelectric fields. We show how to restore
  rotational symmetry for a nonabelian theory with a gauge fixing condition $A^a_3=0$. We also discuss similarities between recovering
  isotropy in the spatial axial gauge and finding gauge invariant quantities in the Weyl gauge in both abelian and nonabelian field theories.

\end{abstract}

\maketitle

\section{Introduction}

Gauge field theories play a very important role in contemporary physics. Their crucial feature is that they can be
expressed in an explicitly Lorentz invariant form,  by introducing additional degrees of freedom. Restriction of a
gauge theory to physically acceptable modes is done by imposing constraints, known as gauge conditions. By
choosing different gauge conditions the theory takes apparently different forms. It is interesting to see how
such different looking quantum field theories theories still represent the same physics.

This paper is concerned with quantum chromodynamics (QCD) in the spatial axial gauge. We show how the apparent anisotropy of this gauge is consistent with the isotropy of the theory. We first recall the analogous situation in a simpler theory -- quantum electrodynamics (QED).
In QED \cite{QED}, the
Hamiltonian and photon propagator  in the Coulomb ($\vec \nabla . \vec A =0$), Lorentz ($\partial_\mu A^\mu =0$), temporal or Weyl
($A_0=0$), and spatial axial gauge ($A_3=0$) differ significantly from each other \cite{Gitman_Tyutin}.
 Nevertheless they describe  the same theory \cite{KH_LL}, \cite{KH-QED-W}. The Hamiltonian in the Coulomb gauge contains a nonlocal
interaction between charge densities $j_0
\frac{1}{\Delta} j_0$, Gauss's law is obeyed by all commutators and operator equations and no further gauge freedom is present.
These properties are also present in the spatial axial gauge with the difference that the Hamiltonian lacks rotational invariance
and the nonlocal interaction has the form $j_0 \frac{1}{\partial_3} j_0$. In the Weyl and Lorentz gauges Gauss's law is not
implemented and must be set as an additional condition on states. There is no nonlocal interaction and the charge density
interacts only through the gauge fields. The gauge conditions are different functionals of the longitudinal and timelike gauge
fields, therefore the charged fields are coupled to "ghost" operators in different combinations.

Not only the Hamiltonians are different: if we analyze particle states, we see that states created by charged particle
creation operators $e^\dagger_s(\vec k)$, $ {\bar e}^\dagger_s(\vec k)$ represent different particles in different
gauges \cite{KH_LL}, \cite{KH-QED-W}. In the Coulomb gauge these states represent electrons and positrons surrounded by an
electric field that
satisfies Gauss's law. The situation is the same in the spatial axial gauge, but the electric field is unphysically
anisotropic. In the Weyl and Lorentz gauges those states represent free particles without any electric or magnetic field.
However, after implementation of Gauss's law in the Weyl and Lorentz gauge and transformation of fields to gauge invariant fields,
or restoration
of rotational symmetry in the spatial axial gauge, particle states describe the same particles in all gauges -- charged
particles surrounded by an isotropic electric field satisfying Gauss's law. The interaction between charged particles mediated
by longitudinal and timelike gauge fields is replaced by a nonlocal interaction between charged densities -- the Coulomb
interaction. In the cases that some interaction between charged particles and photon "ghosts" is left, it can be shown to
have no observable consequences. The time evolution of state vectors in the quotient space of observable particles is
identical in all gauges and so we can say that they manifestly describe the same theory \cite{KH_LL}, \cite{KH-QED-W}.

 It is desirable to establish the same equivalence for QCD, but when we start to deal with a nonabelian theory,
 complications coming from noncommutativity of gauge fields arise.
  We might expect that the Coulomb gauge formulation will play a
  similar role as in the QED case - the formulation expressed completely in terms of isotropic gauge invariant fields. If we
  try to apply the Dirac-Bergmann procedure to quantize QCD (as we can do for QED), the noncommutativity of the gauge fields leads to difficulties. This is the reason why the direct quantization of QCD in the Coulomb gauge has generally been avoided.
  Formulations using different approaches have been given by Schwinger \cite{Schw_QCD}, Gribov
  \cite{Gr} and by Christ and
  Lee \cite{Chr_TDL}.
   In contrast, QCD in the Weyl gauge is relatively easy to quantize \cite{KH_QCD-W1}. It is also possible to follow the
   same procedure as for QED to implement Gauss's law \cite{MB_LC_KH}, find gauge
   invariant fields \cite{LC_MB_KH} and express the Hamiltonian in terms of gauge invariant quantities \cite{MB_LC_KH2}.
The gauge invariant field is transverse, which suggests a direct connection to the Coulomb gauge field \cite{KH_QCD-W2}.
In fact the Hamiltonian expressed in the gauge invariant fields
shows striking similarities to the Hamiltonian in the Coulomb gauge. It was noticed recently that the transverse gauge
invariant chromoelectric field  is not hermitian \cite{KH_HR}. This surprising fact is the cause \cite{KH_HR} of the discrepancies between
the forms of the Hamiltonians published by Schwinger \cite{Schw_QCD}, Christ and Lee  \cite{Chr_TDL}, and Gribov \cite{Gr}. The Hamiltonian also contains a term
analogous to the "ghost" portion of the QED Hamiltonian which in QED has no physical effect. In QCD, this can not be as rigorously
shown, but its physical consequences are limited to radiative corrections \cite{KH}.

   I will consider QED and QCD in the spatial axial gauge and show how to recover rotational invariance. As
  in the case of the Coulomb and Weyl gauges, noncommutativity of the Lie algebra of QCD causes difficulties. Hence it is not
  obvious what the isotropic theory would look like, although the Weyl gauge is very suggestive. I will compare procedures of restoring
  rotational invariance in the spatial axial gauge, and implementing Gauss's law and gauge invariant fields in the Hamiltonian in the Weyl
  gauge. I will show that these two procedures produce identical quantities in QED but not in QCD and the differences will be pointed out.

\section{QED in spatial axial gauge $A_3=0$.}

To understand better features and formalism of QCD quantized in the spatial axial gauge we recall the
simpler QED case \cite{KH_LL}. In \cite{KH_LL} all field operators were given in the momentum space. Here we will use the position
representation which seems to be more transparent for a nonabelian theory.

The Lagrangian has the very well known form:
\be
  \mathcal L  = \psi^{\dagger} \gamma^0 \left( i \gamma^{\mu} \partial_{\mu} - m \right) \psi - \frac{1}{4} F_{IJ}F_{IJ} + \frac{1}{2}
F_{0I}F_{0I} - j_I A_I + j_0 A_0 - A_3 G
\ee
From now we will use a noncovariant notation when vector components are assigned with subscripts: $\vec A =(A_1, A_2, A_3)$.
Indices running through values $1,2,3$ will be assigned with capital letters $I,J,...$ and indices running only through $1,2$
will be assigned with small letters $i,j,...$. The gauge-fixing field $G$ allows us to treat the gauge condition $A_3=0$ as one
of the Euler-Lagrange equations but at the end we have to make sure that our theory is indeed quantum electrodynamics.

  The Euler-Lagrange equations of motion have the form:
      $$ \begin{array}{rcccl}
                 \partial_0 F_{0I} - \partial_J F_{IJ} - j_I + \delta_{I3} G &=& 0& \qquad (i \gamma^\mu \partial_\mu + e\gamma^\mu A_\mu - m)\psi &= 0 \hskip 3cm \\
                 \partial_I F_{0I} + j_0 & =& 0& \qquad \psi^{\dagger}\gamma^0 (i \gamma^\mu \overleftarrow{\partial}_\mu - e\gamma^\mu A_\mu + m) &= 0 \hskip 3cm  \\
                                     A_3 & =& 0&
                                                                                                                       \end{array} $$
 The conjugate momenta are
$\Pi_I = \frac{\delta \mathcal L}{\delta \partial_0 A_I} = - F_{0I}$,
$\Pi_0 = \frac{\delta \mathcal L}{\delta \partial_0 A_0} = 0$,
$\Pi_G = \frac{\delta \mathcal L}{\delta \partial_0 G}= 0$,
$\Pi_{\psi} = \frac{\delta \mathcal L}{\delta \partial_0 \psi} = i \psi^{\dagger}$,
$\Pi_{\psi^\dagger} = \frac{\delta \mathcal L}{\delta \partial_0 \psi^{\dagger}} = 0$.

The theory contains constraints and the Dirac-Bergmann method \cite{D-B} is used to obtain the commutation rules and the Hamiltonian.
This procedure produces secondary and tertiary constraints and among them, the gauge condition, $A_3 \approx 0$, and Gauss's law,
$ \partial_I \Pi_I - j_0  \approx 0$. All constraints are
second-class constraints, which implies that we have no residual gauge freedom and all of them can be considered as operator
equations. The Dirac-Bergman procedure further gives a prescription how to construct Dirac brackets which become (anti-) commutators, and
produces a Hamiltonian which can be represented in terms of independent variables, matter fields $\psi$, $\psi^\dagger$ and gauge fields
$A_i$ and $\Pi_i$ ($i=1,2$), as
\ba
  H &=&  \int \! dx^3 \Big\lgroup \psi^{\dagger}\gamma^0 (- i \vec{\gamma}.\vec{\nabla} + m )\psi +  j_i A_i + \frac{1}{4} F_{ij}F_{ij} +
\frac{1}{2} (\partial_3 A_i)(\partial_3 A_i) + \frac{1}{2} \Pi_i \Pi_i  \nn \\
& & \hskip 1.5cm -\frac{1}{2} (\partial_i \Pi_i - j_0) \frac{1}{(\partial_3)^2} (\partial_i \Pi_i - j_0) \Big\rgroup
\label{H_QED}
\ea
The fields obey standard (anti-)commutation rules:
$$ \begin{array}{rl}
         &\{ \psi(\vec x), \psi^{\dagger} (\vec y) \} =  \delta (\vec x - \vec y)   \\
         &[ A_i(\vec x), \Pi_j (\vec y) ]= i \delta_{ij} \delta (\vec x - \vec y)  \qquad i,j = 1,2
\end{array} $$

 The Hamiltonian (\ref{H_QED}) contains  a nonlocal interaction between charge densities similar to in the Coulomb gauge but spatially anisotropic.
After completing the procedure all the fields are gauge invariant and we eliminated all gauge dependent degrees of freedom.
The gauge-fixing field $G$ is also eliminated from the Hamiltonian and has no consequences on physical variables. The electrons
(represented as states constructed from vacuum by electron field creation operator $e^\dagger$) are surrounded by a severely
anisotropic
electric field that has the first two components vanishing and the third one proportional to an integral of the charge density
$j_0$.
\ba
& \langle e_s(\vec k)|\Pi_i(\vec x)|e_s(\vec k)\rangle = 0 , \qquad i,j=1,2  \nn \\
& \langle e_s(\vec k)|\Pi_3(\vec x)|e_s(\vec k)\rangle =- \langle e_s(\vec k)|\frac{1}{\partial_3}j_0 (\vec x)|e_s(\vec k)\rangle
\ea

This unphysical feature is caused by using common rotationally invariant states $|e_s(\vec k)\rangle$, which are familiar in
other, isotropic, gauges. But if we have such an anisotropic Hamiltonian as (\ref{H_QED}) we can not expect that its eigenstates will
be isotropic. If gauge fixing has no physical consequence and all physical quantities are the same in all gauges,
the appropriate states for calculating expectation values will contain an asymmetry which cancels all
anisotropy from all physical quantities.

 Accordingly,  we can look for a unitary transformation $e^{\Lambda}$, (with $\Lambda^\dagger=-\Lambda$), connecting
 anisotropic and isotropic states
\be
|n_i\rangle_{anisotr.} = e^{-\Lambda} |{\bar n}_i\rangle_{isotr.}
\ee
This transformation, when acting on an operator valued field $O$ in the spatial axial gauge, will produce a
transformed field $\bar O$
\be
\bar O = e^{\Lambda} O e^{-\Lambda}.
\label{transformation}
\ee
For QED, $\Lambda$ has the explicit form \cite{KH_LL}:
\be
  \Lambda = i \int dx^3
  j_0(\vec x)\chi(\vec x) \qquad
   \text{where:}
   \qquad \chi(\vec x)  = \frac{\partial_i}{\Delta}A_i( \vec x)
\label{lambda_QED}
\ee
  and produces the following transformed fields:
\ba
{\bar A}_i &=&A_i \label{A-transformed-QED}\\
{\bar{\Pi}}_I &=& \Pi_I - \delta_{I3}\frac{1}{\partial_3}j_0 + \frac{\partial_I}{\Delta}j_0  \label{Pi-transformed-QED}\\
\bar \psi &=& e^{-i e \chi} \label{psi-transformed-QED}\psi
\ea
 These fields are not necessarily isotropic. For example, $\psi$ was isotropic from the beginning but the anisotropic form of Gauss's law
 in the spatial axial gauge prohibits $\Pi_I$ to be also isotropic. At this point they are just some transformed fields, whose role
 will be clarified later.
 The transformed Hamiltonian $\bar H$ is:
\ba
 \!\!\!\!\!\! \bar H = \!\!\! \int \! dx^3 \Big\lgroup \psi^\dagger \gamma^0 (- i \vec{\gamma}.\vec{\nabla} + m )\psi
  - \vec j . \vec{\nabla}\chi
  +  j_i A_i + \frac{1}{4} F_{ij}F_{ij} + \frac{1}{2} (\partial_3 A_i)(\partial_3 A_i)
  + \frac{1}{2} \bar{\Pi}_I \bar{\Pi}_I  \Big\rgroup
\label{H_QED-isotr1}
\ea

 The gauge conditon $A_3=0$ does not allow any unitary transformation which would produce completely isotropic
$\bar A_I$ but we expect only physical electromagnetic fields to be rotationally invariant.
The Hamiltonian (\ref{H_QED-isotr1}), is already isotropic, even though it
does not appear so. To show its rotational invariance we could
express all fields in terms of creation and anihilation operators as
was done in \cite{KH_LL} or we can transform the gauge fields without changing electric and magnetic fields:
\be
 {\mathcal A}_I = A_I - \partial_I \chi \label{A_field}
\ee
 There are no longitudinal degrees of freedom present and by
introducing the new gauge field we just redistribute the transverse photon
polarization to the three components of the vector $\vec \mathcal A$. It is also convenient to separate the interaction out of the conjugate
momenta $\Pi_I$, and introduce purely transverse electric fields $\mathcal P_I$:
\ba
    &{\mathcal P}_i& = \bar{\Pi}_i - \frac{\partial_i}{\Delta}j_0 = \Pi_i \label{P,A-isotr} \\
    &{\mathcal P}_3& = \bar{\Pi}_3 - \frac{\partial_3}{\Delta} j_0 = \Pi_3 - \frac{1}{\partial_3} j_0
    = - \frac{\partial_i}{\partial_3} \Pi_i \nn
\ea
  The new fields $\mathcal A_I$, $\mathcal P_I$ have the same properties as
 the fields in the Coulomb gauge, they are transverse and have the same commutation relation:
\ba
  &\partial_I {\mathcal A}_I = 0 \nn \\
  &\partial_I {\mathcal P}_I = 0 \nn  \\
  &[{\mathcal A}_I (\vec x), {\mathcal P}_J (\vec y) ] = i
         \left( \delta_{IJ} - \frac{\partial_I \partial_J}{\Delta} \right) \delta(\vec x - \vec y)
\ea
The Hamiltonian expressed in terms of ${\mathcal A}_I$ and ${\mathcal
P}_I$ has the familiar Coulomb gauge form:
\ba
  H =  \int \! dx^3 \Big\lgroup \psi^\dagger \gamma^0 (- i \vec{\gamma}.\vec{\nabla} + m )\psi
  - \vec j . \vec{\nabla}\chi
  +  \vec j . \vec{\mathcal A} + \frac{1}{4} {\mathcal F}_{IJ}{\mathcal
  F}_{IJ}
+ \frac{1}{2} {\mathcal P}_I {\mathcal P}_I - \frac{1}{2}\ j_0 \frac{1}{\Delta} j_0 \Big\rgroup
\label{H_QED_Coul}
\ea
The field strength tensor is ${\mathcal F}_{IJ}=\partial_I {\mathcal
A}_J - \partial_J {\mathcal A}_I$.

In this way, the transformation (\ref{transformation}) brought all terms in the Hamiltonian (\ref{H_QED}) to a rotationally
invariant form and produced the Coulomb nonlocal interaction between charge
densities, and at the same time canceled the earlier nonlocal
unisotropic  interaction. The particles are now accompanied by the Coulomb electric field
\ba
 \langle e_s(\vec k)|\bar \Pi_I(\vec x)|e_s(\vec k)\rangle =- \langle e_s(\vec k)|\frac{\partial_I}{\Delta}j_0 (\vec x)
                                       |e_s(\vec k)\rangle.
\ea
Now we can easily identify rotationally invariant fields $\psi$, $\psi^\dagger$, $\mathcal A_I$ and $\mathcal P_I$ (or $\bar{\Pi}_I$).

  It is interesting to notice formal similarities between the procedure we have just completed and the one performed
  in the Weyl and Lorentz gauge \cite{KH_LL} in order to find gauge invariant fields.
  Formally the transformations have the same form
(\ref{lambda_QED}), despite having a different physical meaning. (In the spatial axial gauge all fields are gauge invariant since
Gauss's law holds as an operator identity which is not violated in any step during the procedure.) Also the transformed charge particle field
(\ref{psi-transformed-QED}) has the same form as the gauge invariant electron field found originally by Dirac \cite{Dirac:1955uv}, and the gauge field
(\ref{A_field}) is transverse and hence has the form of the gauge invariant field. At the end we have the transformation, Hamiltonian and transformed
fields which are identical to the gauge invariant quantities in the Weyl and Lorentz gauge.
It is an intriguing question whether such formal similarities persist also in QCD.

\section{QCD in spatial axial gauge $A^a_3=0$.}

Having recalled the simpler Abelian case, we now apply the same approach to a nonabelian
gauge theory. From the beginning, we have to expect some complications
coming from noncommutativity of the gauge fields.

The Lagrangian of QCD has the following form:
\be
 \mathcal L  = \psi^\dagger \gamma^0 \left( i \gamma^{\mu} \partial_{\mu} + g \gamma^\mu A_\mu^a \frac{\tau^a}{2} - m \right)\psi - \frac{1}{4} F^a_{\mu\nu}
                {F^a}^{ \mu\nu} - {A^a}^{3} G^a
\label{L_QCD}
\ee
 The field-strength tensor is given as $F_{\mu\nu}^a  =   \partial_\mu A_\nu^a - \partial_\nu A^a_\mu + g f^{abc}A^b_\mu
 A_\nu^c$. All gauge fields now have a color index of $SU(3)$ group with structure constants $f^{abc}$, and generators
 $\tau_a$, which obey commutation relations $[\frac{\tau^a}{2} , \frac{\tau^b}{2} ] = i f^{abc} \frac{\tau^c}{2}$.
 In analogy to QED we introduce a set of gauge fixing fields $G^a$. In contrast to the Weyl gauge \cite{KH_QCD-W1}
 these fields do not provide us with canonically conjugate momenta to $A^a_0$, but enable us to treat the gauge
 condition as one of the equations of motion.

  The Euler-Lagrange equations derived from the Lagrangian (\ref{L_QCD}) have the following form:
\ba
    \left(\delta^{ab} \partial_\mu - g f^{abc} A_\mu^c \right) {F^b}^{ \mu\nu} + j^{a \nu} + \delta^\nu_3 G^a &= 0 \nn \\
    \left( i \gamma^\mu \partial_\mu + g \gamma^\mu A_\mu^a \frac{\tau^a}{2}  - m \right)\psi &= 0  \\
    \psi^\dagger \gamma^0 \left(i \gamma^\mu \overleftarrow{\partial}_\mu - g \gamma^\mu A_\mu^a \frac{\tau^a}{2} + m \right) &= 0 \nn \\
                                    A^{a}_3 & = 0 \nn
\label{Eq_of_mot_QCD}
\ea
 where the quark current density is $j^a_\mu = g \psi^\dagger \gamma^0 \gamma_\mu \frac{\tau^a}{2} \psi$.
 For a transition to the Hamiltonian
 formalism we find conjugate momenta:
 $\Pi^a_I = \frac{\delta \mathcal L}{\delta \partial_0 A^a_I} = - F^a_{0I}$,
 $\Pi^a_0 = \frac{\delta \mathcal L}{\delta \partial_0 A^a_0} = 0$,
 $\Pi^a_G = \frac{\delta \mathcal L}{\delta \partial_0 G^a}= 0$,
 $\Pi_{\psi} = \frac{\delta \mathcal L}{\delta \partial_0 \psi} = i \psi^{\dagger}$,
 $\Pi_{\psi^\dagger} = \frac{\delta \mathcal L}{\delta \partial_0 \psi^{\dagger}} = 0$. As before, we are dealing
 with a system with constraints and therefore the Dirac-Bergman procedure \cite{D-B} is used to obtain Hamiltonian and (anti-) commutation relations.
 The Hamiltonian of QCD in the spatial axial gauge is:
\ba
 H_0 &= \int\!\! dx^3 \Big\lgroup \psi^\dagger \gamma^0 (- i \vec{\gamma}.\vec{\nabla} + m )\psi + \frac{1}{2} \Pi^a_i \Pi^a_i - \frac{1}{2}
      (\partial_i \Pi^a_i) \frac{1}{(\partial_3)^2} (\partial_j\Pi^a_j)+\nn \\
      & \qquad +\frac{1}{2} (\partial_3 A^a_i)(\partial_3 A^a_i)
      + \frac{1}{2} (\partial_i A^a_j)(\partial_i A^a_j)  - \frac{1}{2} (\partial_i A^a_j)(\partial_j A^a_i) \Big\rgroup \nn  \\
 H_g  &= \int\!\! dx^3 \Big\lgroup  j^a_i A^a_i - gf^{abc}(\partial_i A^a_j) A^b_i A^c_j + (\partial_i \Pi^a_i)
         \frac{1}{(\partial_3)^2} (j^a_0 + J^a_0) \Big\rgroup \label{H_QCD_SAG} \\
 H_{g^2}  &= \int\!\! dx^3 \Big\lgroup \frac{1}{4} g^2 f^{abc}f^{ade} A^b_i A^c_j A^d_i A^e_j  - \frac{1}{2} (j^a_0 + J^a_0)
         \frac{1}{(\partial_3)^2} (j^a_0 + J^a_0)  \Big\rgroup \nn
\ea
For later convenience we have split the Hamiltonian into three terms with different powers of the coupling constant $g$. We again use
a noncovariant notation $\vec A^a =(A^a_1, A^a_2, A^a_3)$ with the same use of indices $ I,J = 1,2,3$, and $i,j=1,2$. The color
charge density is given by $J^a_0 = g f^{abc} A_i^b \Pi_i^c$.   The
Hamiltonian (\ref{H_QCD_SAG}), containing only independent degrees of freedom $\psi$, $\psi^\dagger$, $A^a_i$ and
$\Pi^a_i$ obeying canonical (anti-)commutation relations
\ba
&\{ \psi(\vec x), \psi^{\dagger} (\vec y) \} =  \delta (\vec x - \vec y) \nn  \\
&[ A^a_i(\vec x), \Pi^b_j (\vec y) ]= i \delta_{ij} \delta^{ab} \delta (\vec x - \vec y),
\ea
is again severely anisotropic.

  We saw in the QED case that even the transformed Hamiltonian (\ref{H_QED-isotr1})
did not look isotropic until we introduced fields (\ref{A_field}) and (\ref{P,A-isotr}). We expect that
similar isotropic, "Coulomb gauge" fields ${\mathcal A}_I$ and ${\mathcal P}_I$ can be used also for a nonabelian theory.
\ba
   {\mathcal A}^a_I &=& A^a_I - \partial_I \chi^a  \nn \\
   {\mathcal P}^a_i &=& \Pi^a_i   \label{P,A-isotr_QCD} \\
   {\mathcal P}^a_3 &=& -\frac{\partial_i}{\partial_3} \Pi^a_i = \Pi^a_3 - \frac{1}{\partial_3} (j^a_0 + J^a_0) \nn
\ea
where $\chi^a = \frac{\partial_i}{\Delta}A^a_i$.  The fields now acquire a color index and
${\mathcal P}^a_3$ contains also a gluon charge
term in such a form that ${\mathcal P}^a_I$ obeys the  "free Gauss law" $\partial_I {\mathcal P}^a_I = 0$.
The fields are transverse and have the same commutation relations as before:
\be
 \partial_I {\mathcal A}^a_I = 0 \qquad \partial_I {\mathcal P}^a_I = 0
\ee
\be
[{\mathcal A}^a_I (\vec x), {\mathcal P}^b_J (\vec y) ] = i \delta^{ab} \left( \delta_{IJ} - \frac{\partial_I \partial_J}{\Delta}
  \right) \delta(\vec x - \vec y)
\label{comm_relations-QCD}
\ee

 To obtain an isotropic Hamiltonian $\bar H$ we have to find a unitary transformation $e^\Lambda$ such that:
\be
\bar H = e^\Lambda H e^{- \Lambda} = H + [\Lambda, H] + \frac{1}{2}[\Lambda, [\Lambda,H]] + ...
\ee
The convenience of using the Baker-Campbell-Hausdorff formula comes from expansions of $\Lambda$, $H$ and $\bar H$ as
power series in the coupling constant.
\ba
 \Lambda &=& \Lambda_g + \Lambda_{g^2} + \Lambda_{g^3} + ... \nn  \\
     H &=& H_0 + H_g + H_{g^2}  \label{expansion} \\
     \bar H &=& {\bar H}_0 + {\bar H}_g + {\bar H}_{g^2}+... \nn
\ea
The expansion of $\bar H$ does not have to terminate and the isotropic Hamiltonian will in general contain all powers of $g$,
similar to what happens in the Weyl gauge when the Hamiltonian is expressed in terms of gauge invariant fields \cite{KH_HR}.
From (\ref{expansion}) we can find
the first few terms of the transformed Hamiltonian
\ba
{\bar H}_0 &=& H_0  \nn \\
{\bar H}_g &=& H_g + [\Lambda_g, H_0] \label{expansion_H} \\
{\bar H}_{g^2} &=& H_{g^2} + [\Lambda_g, H_g] + [\Lambda_{g^2}, H_0] +\frac{1}{2} [\Lambda_g,[\Lambda_g, H_0]] \nn
\ea
This pattern suggests a way of finding the transformation $e^\Lambda$ order by order. If we know $\Lambda_{g^{n-1}}$,
we can calculate all its commutators and if we further assume some isotropic form of ${\bar H}_{g^n}$, we can find
$[\Lambda_{g^n}, H_0]$ and make a well-motivated guess for the form of $\Lambda_{g^n}$.

 The free Hamiltonian is isotropic from the beginning. This fact we already used in (\ref{expansion})
 where we assumed that the lowest order of
$\Lambda$ is proportional to the first power of the coupling constant $g$.
As in the QED case, $\bar H_0$ can be written in terms of ${\mathcal
 A}^a_I$ and ${\mathcal P}^a_I$ to see its rotational invariance explicitly:
\be
 \bar H_0 = \int\!\! dx^3 \Big\lgroup \psi^\dagger \gamma^0 (- i \vec{\gamma}.\vec{\nabla} + m )\psi + \frac{1}{2} {\mathcal P}^a_I
 {\mathcal P}^a_I + \frac{1}{4} (\partial_I {\mathcal A}^a_J - \partial_J {\mathcal A}^a_I)(\partial_I {\mathcal A}^a_J -
 \partial_J {\mathcal A}^a_I)  \Big\rgroup \label{H_0_isotropic}
\ee
The free Hamiltonian (\ref{H_0_isotropic}) is identical to the free Hamiltonian in the Coulomb gauge.

To find $\Lambda_g$ we can assume $\bar H_g$ to be of the form:
\be
{\bar H}_g = \int\!\! dx^3 \Big\lgroup  j^a_I {\mathcal A}^a_I - gf^{abc}(\partial_I {\mathcal A}^a_J) {\mathcal A}^b_I
{\mathcal A}^c_J  \Big\rgroup \label{H_g_isotropic}
\ee
The Hamiltonian of the spatial axial gauge (\ref{H_QCD_SAG}) contains no $A_3^a$.
To recover the third component of the gauge field ${\mathcal A}^a_3 = - \partial_3 \chi^a$ we make use of the commutation
relation between $i \int dx^3 \ \chi^a  j^a_0$ and the kinetic energy of the spinor field:
\be
\Big[ i \int\!\! dx^3 \ \chi^a(\vec x) j^a_0(\vec x), \int\!\! dy^3 \ \psi^\dagger(\vec y) \gamma^0 \left(- i \vec{\gamma}.\vec{\nabla} + m
\right) \psi(\vec y)\Big] = - \int\!\! dx^3 \ {\vec j}^a.\vec{\nabla}\chi^a \label{commutator-1}
\ee
In the case of QED there was no interaction among gauge fields and $i \int dx^3 \ \chi  j_0$ was all we needed to
find the desired transformation. In QCD, such a simple transformation does not produce a completely satisfactory result.
It is natural  to alter this term by adding the
gluon charge density $J^a_0$ to the quark charge density $j^a_0$, to obtain the second term in (\ref{H_g_isotropic}).
Because the gluon charge density $J^a_0$ does not
contain third components of $\mathcal A^a_I$ and $\Pi^a_I$ we have to include additional terms. Then $\Lambda_g$ will have
the following form:
\ba
\Lambda_g &=& i \int dx^3 \Big\lgroup \chi^a \left(j^a_0 + J^a_0 \right)- \frac{1}{2}g f^{abc} \left[ \chi^a
                        \left(\partial_i \chi^b \right) \Pi^c_i - \chi^a \left(\partial_3 \chi^b \right)
                        \frac{\partial_i}{\partial_3}\Pi^c_i \right] \Big\rgroup  \nn \\
           &=& i \int dx^3 \ \chi^a \Big\lgroup j^a_0 + \frac{1}{2} \left( J^a_0 + {\mathcal J}^a_0 \right)
                         \Big\rgroup \label{Lambda_g}
\ea
where ${\mathcal J}^a_0 = g f^{abc} {\mathcal A}_I^b {\mathcal P}_I^c $.

The next order of $\bar H$ is given by the third equation of (\ref{expansion_H}). The commutators of
$\Lambda_{g}$ produce terms proportional to $\vec j^a$. Such terms are not expected to be in the isotropic Hamiltonian
of order $g^2$ and therefore must be canceled by $[\Lambda_{g^2}, H_0]$. Using the commutator
\be
\Big[ j^a_0(\vec x), \int\!\! dy^3 \ \psi^\dagger(\vec y) \gamma^0 \left(- i \vec{\gamma}.\vec{\nabla} + m
\right) \psi(\vec y)\Big] = - i \ \vec{\nabla}.{\vec j}^a (\vec x) \label{commutator-2}
\ee
we can find an analog to (\ref{commutator-1}), and
hence find the part of $\Lambda_{g^2}$ proportional to $j^a_0$ which produces such terms. Following the same way as before we
can assume that quark and gluon charge density $j^a_0$ and $J^a_0$ play a similar role in $\Lambda_{g^2}$ and then find the
correcting terms with the right order  of operators. $\Lambda_{g^2}$ will be of the form:
\ba
\Lambda_{g^2} = -\frac{i}{4} \int\!\! dx^3 \!\!\! &\Big\lgroup g \left( \frac{\partial_I}{\Delta} \Psi^a_{(1)I} \right) \left( j^a_0 +
   {\mathcal J}^a_0 \right) + g \left( j^a_0 + {\mathcal J}^a_0 \right) \left( \frac{\partial_I}{\Delta} \Psi^a_{(1)I} \right) \nn \\
    &+ \ g \left( \frac{1}{\partial_3} \Psi^a_{(1)3} \right) \left( j^a_0 + J^a_0 \right) + g \left( j^a_0 + J^a_0 \right)
    \left( \frac{1}{\partial_3} \Psi^a_{(1)3} \right) \nn \\
    &- \frac{1}{6} \ g^2 f^{abc}f^{ade} \left\{ \chi^b (\partial_I \chi^c) \chi^d {\mathcal P}^e_I + {\mathcal P}^e_I \chi^d
    (\partial_I \chi^c) \chi^b \right\}  \Big\rgroup \label{Lambda_g2}
\ea
where $ \Psi^a_{(1)I} = f^{abc} \chi^b \left( A^c_I - \frac{1}{2} \partial_I \chi^c \right) $. The same
$\Psi^a_{(1)I}$ is a part of so called "resolvent field" needed to implement Gauss's law in QCD in the temporal gauge and
find gauge invariant operators \cite{MB_LC_KH, LC_MB_KH}.

The next order of the isotropic Hamiltonian $H_{g^2}$ obtained by using (\ref{Lambda_g2}) is
\ba
   {\bar H}_{g^2} = \int\!\! dx^3 \Big\lgroup \frac{1}{4} g^2 f^{abc}f^{ade} {\mathcal A}^b_I {\mathcal A}^c_J
   {\mathcal A}^d_I {\mathcal A}^e_J  - \frac{1}{2} (j^a_0 + {\mathcal J}^a_0) \frac{1}{\Delta} (j^a_0 + {\mathcal J}^a_0)
   \Big\rgroup + \textrm{{\small const.}} \label{H_g2_isotropic}
\ea
where "const" refers to the following $\mathcal C$-number function:
\be
  \textrm{\small const.} = \int\!\! dx^3 \frac{1}{8} g f^{abc} \left\{ \left[ ( {\mathcal A}^b_I - A^b_I ), [{\mathcal P}^a_I, \frac{1}{\Delta} {\mathcal J}^c_0] \right] +
  \left[\chi^b, [{\mathcal P}^a_3, \frac{1}{\partial_3}J^c_a] \right]  \right\} \label{constant}
\ee
This constant term comes from different operator ordering in (\ref{Lambda_g2}). After evaluation of the commutators, (\ref{constant}) does not contain any
operators, producing just a $\mathcal C$-number function. This constant causes an unobservable shift of the ground state energy and therefore has no
physical consequences.

Therefore, the isotropic Hamiltonian up to the second order in the coupling constant has
the following form:
\be
 \bar H = \int\!\! dx^3 \Big\lgroup \psi^\dagger \gamma^0 (- i \vec{\gamma}.\vec{\nabla} + m )\psi + j^a_I {\mathcal A}^a_I +
                 \frac{1}{2} {\mathcal P}^a_I
                {\mathcal P}^a_I + \frac{1}{4} {\mathcal F}^a_{IJ} {\mathcal F}^a_{IJ} - \frac{1}{2} (j^a_0 +
                {\mathcal J}^a_0) \frac{1}{\Delta} (j^a_0 + {\mathcal J}^a_0)  \Big\rgroup + \textrm{\small const.}
 \label{H_QCD_0_g_g2_isotr}
\ee
where: ${\mathcal F}_{IJ}^a = \partial_I {\mathcal A}_J - \partial_J {\mathcal A}^a_I + g f^{abc}{\mathcal A}^b_I {\mathcal A}^c_J $
The Hamiltonian (\ref{H_QCD_0_g_g2_isotr}) has the same form as the Hamiltonian in the Coulomb gauge \cite{Schw_QCD,Gr,Chr_TDL,KH_HR}.

 The next order, $\bar H_{g^3}$, can be obtained from
\ba
    {\bar H}_{g^3} &=& \ [ \Lambda_{g^3}, H_0] + [ \Lambda_{g^2}, H_g] + [ \Lambda_g, H_{g^2}] \nn \\
                   &+& \frac{1}{2} [ \Lambda_g, [ \Lambda_g, H_g]] +  \frac{1}{2} [ \Lambda_g, [ \Lambda_{g^2}, H_0]] +
                   \frac{1}{2} [ \Lambda_{g^2}, [ \Lambda_g, H_0]]  \label{H_g3_isotropic-1} \\
                   &+& \frac{1}{3!} [ \Lambda_g, [ \Lambda_g,  [ \Lambda_g, H_0]]] \nn
\ea
The calculations follow the previous pattern. The known
commutators produce terms proportional to $j^a_I$ which must be canceled by $[\Lambda_{g^3}, H_0]$. Therefore,
$\Lambda_{g^3}$ must be of the form $ \int dx^3 f^a(A) j^a_0 $, where we find $f^a(A)$ to be an operator
function of the gauge field $A^a_i$:
\ba
f^a(A)&=& \frac{\imath}{4 !} \ g^2 \ f^{abc} \ \frac{\partial_I}{\Delta} \Bigg\{ \chi^b \left( 2\ \Psi^c_{(1)I} +
11\ \frac{\partial_I \partial_J}{\Delta} \Psi^c_{(1)J} - 25\ \frac{\partial_I}{\Delta} \Psi^c_{(1)3} \right) \nn \\
&-&(\partial_I \chi^b) \left( 21\ \frac{\partial_J}{\Delta} \Psi^c_{(1)J}- 39\ \frac{1}{\partial_3} \Psi^c_{(1)3} \right)
- 4\ A^b_I \left( 2\ \frac{\partial_J}{\Delta} \Psi^c_{(1)J}+ \frac{1}{\partial_3} \Psi^c_{(1)3} \right)\nn \\
&+&\! 4 !\ f^{cde} \! \Bigg[ \chi^b(\partial_I \chi^d)\chi^e - \frac{1}{8} \frac{1}{\partial_3}
\Big( \chi^b(\partial_I \chi^d)(\partial_3\chi^e) \Big) + \frac{1}{\partial_3}
\Big( (\partial_3 \chi^b)\chi^d \Big) A_I^e - \chi^b \frac{1}{\partial_3} \Big(\chi^d (\partial_3 A_I^e) \Big) \nn \\
&-& \frac{1}{\partial_3} \Bigg( (\partial_3 \chi^b) \frac{1}{\partial_3} \Big[(\partial_3 \chi^d)(A_I^e-
\partial_I \chi^e) \Big] +  \frac{1}{\partial_3} \Big[ \chi^b (\partial_3 \chi^d) \Big](\partial_3 A_I^e)\Bigg) \Bigg]
\Bigg\} \label{Lambda_f_g3} \\
&+& \frac{1}{12}\ g^2\ f^{abc}\ \frac{1}{\partial_3} \Bigg\{2\ (\partial_3 \chi^b) \left(\frac{\partial_I}{\Delta}
\Psi^c_{(1)I} \right) - 5\ (\partial_3 \chi^b) \left(\frac{1}{\partial_3}
\Psi^c_{(1)3} \right) \Bigg\} \nn
\ea
Recall that $ \Psi^a_{(1)I} = f^{abc} \chi^b \left( A^c_I - \frac{1}{2} \partial_I \chi^c \right) $, as above.

The next step would be to include the gluon charge density $J^a_0$ in the same combination as the quark charge
density $j^a_0$ and
find terms with different operator order to recover rotational invariance. We can assume some
isotropic form of $\bar H_{g^3}$. The
QED case and the previous orders of $\bar H$ suggest the Coulomb gauge form \cite{Schw_QCD, Gr, Chr_TDL, KH_HR}:
\be
 {\bar H}_{g^3} = -gf^{abc} \!\!\! \int\!\! dx^3 dy^3 \left( j^a_0 + {\mathcal J}^a_0 \right) (\vec x)
                \frac{1}{\Delta}\left( {\mathcal A}^b_I(\vec x) \frac{\partial_I}{\Delta} \delta(\vec x -\vec y) \right)
                \left( j^c_0 + {\mathcal J}^c_0 \right) (\vec y)
\ee

  In this way one can continue in finding higher and higher orders of the transformation and recover rotational invariance.

  Without calculating the higher orders we can already compare the isotropy restoring transformation (\ref{Lambda_g}), (\ref{Lambda_g2}) with the
transformation (3.5) in \cite{LC_MB_KH} used to find gauge invariant fields in the Weyl gauge. The first orders of this transformation are:
\be
  \mathcal U_\mathcal C = \exp \left\{ \imath \int\!\! dx^3\ \left[ \chi^a
                           - g \left( \frac{\partial_I}{\Delta}\Psi^a_{(1)I} \right)  \right] j_0^a  \right\} + O(g^3)
  \label{U_C}
\ee
where $ \Psi^a_{(1)I} = f^{abc} \chi^b \left( A^c_I - \frac{1}{2} \partial_I \chi^c \right) $.
 We can see that
there are significant differences between these two transformations. The transformation (\ref{U_C}) does not contain any conjugate momenta
$\Pi^a_I$. Terms proportional to $\Pi^a_I$ are necessary in (\ref{Lambda_g}) and (\ref{Lambda_g2}) to make the parts of the Hamiltonian
(\ref{H_QCD_SAG}) containing the gluon charge density $J^a_0$ rotationally symmetric. There are some similarities between the part
of the transformations proportional to the spinor charge density $j^a_0$. They are identical in the first order
of the coupling constant $g$. The term
$\int dx^3  g \left( \frac{1}{\partial_3} \Psi^a_{(1)3} \right) \left( j^a_0 + J^a_0 \right)$ in (\ref{Lambda_g2}) can be written using
integration by parts and Gauss's law as
\begin{displaymath}
g \int dx^3 \Big[ \left( \frac{\partial_3}{\Delta} \Psi^a_{(1)3} \right) \left( j^a_0 + J^a_0 \right) + \Psi^a_{(1)3} \mathcal P^a_3 \Big]
\end{displaymath}
 In this way the part of (\ref{Lambda_g2}) proportional to $j^a_0$ resembles the relevant order of the resolvent field but in an unisotropic way:
\ba
 -\ \imath \ \frac{g}{2} \int\!\! dx^3 \! \left[ \left( \frac{\partial_i}{\Delta} \Psi^a_{(1)i} \right)
     +  2 \left( \frac{\partial_3}{\Delta} \Psi^a_{(1)3} \right)   \right] j^a_0
\ea

 If we disregard nonabelian terms in both transformations, all the
 differences go away and we recover QED.

\section{Transformed fields}

 We already compared the isotropy restoring transformation (\ref{Lambda_g}), (\ref{Lambda_g2}) in the spatial axial gauge and the transformation
(\ref{U_C}) creating gauge invariant fields in the Weyl gauge.
  To complete the discussion we will look at the transformed fields. These fields do not have to be necessarily isotropic, as we could already see
in the QED case. The isotropic fields are those which appear in the isotropic Hamiltonian and have isotropic commutation relations $\psi$,
$\psi^{\dagger}$, $\mathcal A^a_I$ and  $\mathcal P^a_I$.

 The isotropy restoring transformation (\ref{Lambda_g}), (\ref{Lambda_g2}) produces the following fields:
\ba
\bar \psi &=& \left( 1 - \imath g \chi^a \frac{\tau^a}{2} + \frac{1}{2} \left( - \imath g \chi^a \frac{\tau^a}{2}
            \right)^2 + \imath g^2 \left( \frac{1}{\partial_3} \Psi^a_{(1)3} \right) \frac{\tau^a}{2} + ... \right)
            \psi \label{psi-transformed-QCD}\\
\bar A^a_I &=& A^a_I + g f^{abc} \left( \chi^b A^c_I - \frac{1}{\partial_3}\left[\left(\partial_3
             \chi^b\right)\left(\partial_I \chi^c \right)\right] \right) \\
             &+& g^2 f^{abc}f^{cde}
             \left( \frac{1}{\partial_3}\left[\chi^b \left(\partial_3 \chi^d \right)\right] A^e_I +
                    \frac{1}{\partial_3}\left[\chi^b \left(\partial_I \chi^d \right)\left( \partial_3 \chi^e
                    \right)\right] \right) + ... \nn \\
\bar \Pi^a_i &=& \Pi^a_i + \frac{\partial_i}{\Delta} \left( j^a_0 + {\mathcal J}^a_0 \right) + g f^{abc} \chi^b \Pi^c_i \\
             &+& \! \frac{1}{2}  g f^{abc} \Bigg[
                  \frac{\partial_i \partial_I}{\Delta} \! \left( \! {\mathcal A}^b_I \frac{1}{\Delta} \left(j^c_0 \! + \!
                  {\mathcal J}^c_0 \right)  +  \frac{1}{\Delta}\left( j^c_0 \! + \! {\mathcal J}^c_0 \right) {\mathcal A}^b_I
                  \right)
                   +  \left( \! \chi^b \frac{\partial_i}{\Delta} \left( j^c_0 \! + \! {\mathcal J}^c_0 \right)
                   + \frac{\partial_i}{\Delta} \left( j^c_0 \! + \! {\mathcal J}^c_0 \right) \chi^b \right) \nn \\
                  &+&  g f^{cde}
                  \left( \frac{1}{\partial_3}\left[\chi^b \left(\partial_3 \chi^d \right)\right] \Pi^e_i +
                          \Pi^e_i \frac{1}{\partial_3}\left[\chi^b \left(\partial_3 \chi^d \right)\right] \right)
                          \Bigg] + ...\nn
\ea
The bar denotes the transformed fields.

  The most interesting is the spinor field $\psi$. In the Abelian theory the form of the transformed isotropic field (\ref{psi-transformed-QED})
  was identical to the gauge invariant field in the Weyl gauge \cite{KH_LL, KH-QED-W, Dirac:1955uv}.
   In the Weyl gauge QCD, the gauge invariant field (4.4) in \cite{LC_MB_KH} has the form:
\ba
\psi_{\textrm{\tiny GI}}&=&
  \exp \left\{\imath g \overline{\mathcal Y^a}\ \frac{\tau^a}{2}\right\} \exp \left\{ \imath g \chi^a \ \frac{\tau^a}{2}\right\} \psi
  \label{psi_GI-QCD} \\
  &=& \left( 1 + \imath g \chi^a \frac{\tau^a}{2} + \frac{1}{2} \left( \imath g \chi^a \frac{\tau^a}{2}
            \right)^2 - \imath g^2 \left( \frac{\partial_I}{\Delta} \Psi^a_{(1)I} \right) \frac{\tau^a}{2} + ... \right) \psi \nn
\ea
 where
 $\overline{\mathcal Y^a}=\frac{\partial_I}{\Delta}\overline{\mathcal A}^a_I$, and $\overline{\mathcal A}^a_I$ is so called "resolvent field"
 needed to implement Gauss's law and find gauge invariant fields in \cite{MB_LC_KH, LC_MB_KH}. The resolvent field can be written as a power series in $g$, and the
 earlier defined $\Psi^a_{(1)I}$ is its first term, proportional to the first power of $g$.
 To avoid confusion we have to point out that $\psi$ in (\ref{psi-transformed-QCD}) and in (\ref{psi_GI-QCD}) are not the same fields.
 In (\ref{psi_GI-QCD}) $\psi$ is the spinor field in the Weyl gauge (not gauge invariant) and $\psi$ in (\ref{psi-transformed-QCD})
is the isotropic spinor field in the spatial axial gauge.
 As we mentioned earlier, $\psi$, and {\it not} the transformed field $\bar \psi$, is rotationally invariant and corresponds to the spinor
field in the Coulomb gauge, similarly as $\psi_{GI}$ in the Weyl gauge. That is the reason for the opposite sign
in (\ref{psi-transformed-QCD}) and (\ref{psi_GI-QCD}).
In (\ref{psi-transformed-QCD}) and (\ref{psi_GI-QCD}) we can clearly recognize an identical part which adds to a separate exponential,
but the rest differs. The form of $\bar \psi$ (\ref{psi-transformed-QCD}) suggests that it could be written similarly to (\ref{psi_GI-QCD}), as
two exponential terms. The first one, common for both (\ref{psi-transformed-QCD}) and (\ref{psi_GI-QCD}),
$\exp \left\{ \imath g \chi^a \ \frac{\tau^a}{2}\right\}$, and the second one in which (\ref{psi-transformed-QCD}) and (\ref{psi_GI-QCD}) differ.

The gauge invariant gauge field in the Weyl gauge, given by the transverse part of the gauge and resolvent fields:
\be
{A^a_{GI}}_I = \left( \delta_{IJ} - \frac{\partial_I \partial_J}{\Delta} \right) \left( A^a_J + \overline{\mathcal A}^a_J \right),
\ee
and its transverse ("hermitianized") canonical momentum [denoted $\mathcal P^{a \textsf{\scriptsize{T}}}_I$ in \cite{KH_HR}]  correspond in the
spatial axial gauge to $\mathcal A^a_I$ and $\mathcal P^a_I$. They are hermitian, obey the same commutation relations (\ref{comm_relations-QCD})
and appear in
the Coulomb gauge Hamiltonian (\ref{H_QCD_0_g_g2_isotr}) in the same way. Thus we can identify them with the Coulomb gauge fields.

\section{Conclusions}

This paper proposes an iterative way of finding the isotropy restoring transformation for QCD in the spatial axial gauge.
By following this procedure one can find the
 transformation as a power series in the coupling constant. I found its first two orders and confirmed that the rotationally invariant
 Hamiltonian is identical, up to a constant, to the Coulomb gauge Hamiltonian in the first two orders of $g$.
 I also compared restoring of isotropy in the spatial axial gauge with "gauge invariance restoration" in the Weyl gauge
 \cite{MB_LC_KH, LC_MB_KH, MB_LC_KH2, KH_QCD-W2, KH_HR, KH}. I found out that the formal
 similarities of these two procedures present in an Abelian gauge theory disappear in the case of noncommutative gauge fields.

  The isotropic fields in the spatial axial gauge were identified with the gauge invariant fields in the Weyl gauge.
In \cite{KH_HR, KH} two forms of the gauge invariant Hamiltonian are presented. The first one uses a nonhermitian gauge invariant
chromoelectric field ${\Pi^{a \textsf{\scriptsize{T}}}_{\text{\scriptsize GI}}}_I$, and corresponds to the one obtained by Gribov  \cite{Gr}. The second one containing "hermitianized"
momenta $\mathcal P^{a \textsf{\scriptsize{T}}}_I$ corresponds to Schwinger's \cite{Schw_QCD} and Christ and Lee's form \cite{Chr_TDL}.
They are identical up to $g^3$ order, but differ in the higher orders of the coupling constant.
Although the first two orders make no distinction between the two Hamiltonians, the presence of the hermitian canonical momenta obeying the
same commutation relations as the chromoelectric fields $\mathcal P^{a \textsf{\scriptsize{T}}}_I$ in \cite{KH_HR, KH} strongly
suggests that the isotropic Hamiltonian will follow Schwinger's and Christ and Lee's form \cite{Schw_QCD, Chr_TDL} in higher
orders of the coupling constant $g$.

\section*{Acknowledgments}

 I would like to thank Gerald Dunne for very useful discussions, and the US DOE for support through the grant DE-FG02-92ER40716.

\end{document}